\documentstyle[12pt]{article}

\catcode`@=11
\def\chkspace{%
  \relax   
  \begingroup\ifhmode\aftergroup\dochksp@ce\fi\endgroup}
\def\dochksp@ce{%
  \unskip              
  \futurelet\chkspct@k\d@chkspc  
}
\def\d@chkspc{%
  \let\nxtsp@ce=\relax
  \ifx\chkspct@k.\else     
    \ifx\chkspct@k,\else
      \ifx\chkspct@k;\else
        \ifx\chkspct@k!\else
          \ifx\chkspct@k?\else
            \ifx\chkspct@k:\else
              \ifx\chkspct@k)\else
              \ifx\chkspct@k(\else
                \ifx\chkspct@k]\else
                  \ifx\chkspct@k-\else
                    \ifx\chkspct@k\egroup\else  
                      \let\nxtsp@ce=\put@space  
                    \fi
                  \fi
                \fi
              \fi
              \fi
            \fi
          \fi
        \fi
      \fi
    \fi
  \fi
  \nxtsp@ce
}
\def\put@space{$\;$}
\catcode`@=12

\def\ra{\relax\ifmmode \rightarrow\else{{$\rightarrow$}}\fi\chkspace}
\def\etal{{\it et al.}\chkspace}

\def\ie{{\it i.e.}\chkspace}

\def\eg{{\it eg.}\chkspace}

\def\ep{{e$^+$e$^-$}\chkspace}
\def\epa{{e$^+$e$^-$ annihilation}\chkspace}

\def\gluino{\relax\ifmmode \tilde{g} \else $\tilde{g}$ \fi\chkspace}

\def\qq{\relax\ifmmode q\overline{q}
\else $q\overline{q}$ \fi\chkspace}
\def\ff{\relax\ifmmode f\overline{f}
\else $f\overline{f}$ \fi\chkspace}

\def\bb{\relax\ifmmode b\bar{b}
       \else $b\bar{b}$ \fi\chkspace}
\def\ccrm{\relax\ifmmode {\rm c}\bar{\rm c}
       \else ${\rm c}\bar{\rm c}$ \fi\chkspace}

\def\tt{\relax\ifmmode {\rm t}\bar{\rm t}
       \else ${\rm t}\bar{\rm t}$ \fi\chkspace}
\def\ss{\relax\ifmmode {\rm s}\bar{\rm s}
       \else ${\rm s}\bar{\rm s}$ \fi\chkspace}
\def\uu{\relax\ifmmode {\rm u}\bar{\rm u}
       \else ${\rm u}\bar{\rm u}$ \fi\chkspace}
\def\dd{\relax\ifmmode {\rm d}\bar{\rm d}
       \else ${\rm d}\bar{\rm d}$ \fi\chkspace}

\def\qqg{\relax\ifmmode q\overline{q}g
\else $q\overline{q}g$ \fi\chkspace}
\def\bbg{\relax\ifmmode b\overline{b}g
\else $b\overline{b}g$ \fi\chkspace}

\def\ttg{\relax\ifmmode t\overline{t}g
\else $t\overline{t}g$ \fi\chkspace}

\def\afb{\relax\ifmmode A_{FB} \else
{{$A_{FB}$}}\fi\chkspace}
\def\afbb{\relax\ifmmode A_{FB}^b \else
{{$A_{FB}^b$}}\fi\chkspace}
\def\pafb{\relax\ifmmode \tilde{A}_{FB} \else
{{$\tilde{A}_{FB}$}}\fi\chkspace}
\def\pafbb{\relax\ifmmode \tilde{A}_{FB}^b \else
{{$\tilde{A}_{FB}^b$}}\fi\chkspace}

\def\pafbzo{\relax\ifmmode \tilde{A}_{FB}|_{O(0)} \else
{{$\tilde{A}_{FB}|_{O(0)}$}}\fi\chkspace}
\def\pafbfo{\relax\ifmmode \tilde{A}_{FB}|_{\oalp} \else
{{$\tilde{A}_{FB}|_{\oalp}$}}\fi\chkspace}
\def\pafbso{\relax\ifmmode \tilde{A}_{FB}|_{\oalpsq} \else
{{$\tilde{A}_{FB}|_{\oalpsq}$}}\fi\chkspace}
\def\pafbto{\relax\ifmmode \tilde{A}_{FB}|_{\oalpc} \else
{{$\tilde{A}_{FB}|_{\oalpc}$}}\fi\chkspace}

\def\pafbbzo{\relax\ifmmode \tilde{A}_{FB}^b|_{O(0)} \else
{{$\tilde{A}_{FB}^b|_{O(0)}$}}\fi\chkspace}
\def\pafbbfo{\relax\ifmmode \tilde{A}_{FB}^b|_{\oalp} \else
{{$\tilde{A}_{FB}^b|_{\oalp}$}}\fi\chkspace}
\def\pafbbso{\relax\ifmmode \tilde{A}_{FB}^b|_{\oalpsq} \else
{{$\tilde{A}_{FB}^b|_{\oalpsq}$}}\fi\chkspace}
\def\pafbbto{\relax\ifmmode \tilde{A}_{FB}^b|_{\oalpc} \else
{{$\tilde{A}_{FB}^b|_{\oalpc}$}}\fi\chkspace}

\def\afbo0{\tilde{A}_{FB}|_{O(0)}}
\def\afbo1{\tilde{A}_{FB}|_{\oalp}}
\def\afbo2{\tilde{A}_{FB}|_{\oalpsq}}
\def\afbo3{\tilde{A}_{FB}|_{\oalpc}}

\def\lam{\relax\ifmmode \Lambda_{\overline{MS}}
       \else {{$\Lambda_{\overline{MS}}$}}\fi\chkspace}
\def\lamuds{\relax\ifmmode \Lambda^{(3)}_{\overline{MS}}
       \else {{$\Lambda^{(3)}_{\overline{MS}}$}}\fi\chkspace}
\def\lamudsc{\relax\ifmmode \Lambda^{(4)}_{\overline{MS}}
       \else $\Lambda^{(4)}_{\overline{MS}}$\fi\chkspace}
\def\lamudscb{\relax\ifmmode \Lambda^{(5)}_{\overline{MS}}
       \else $\Lambda^{(5)}_{\overline{MS}}$\fi\chkspace}

\def\alp{\relax\ifmmode \alpha_s\else $\alpha_s$\fi\chkspace}
\def\alpbar{\relax\ifmmode \bar{\alpha_s}
       \else $\bar{\alpha_s}$\fi\chkspace}
\def\alpmz{\relax\ifmmode \alpha_s(M_Z)\else $\alpha_s(M_Z)$\fi\chkspace}
\def\alpmzsq{\relax\ifmmode \alpha_s(M_Z^2)
       \else $\alpha_s(M_Z^2)$\fi\chkspace}

\def\oalp{\relax\ifmmode O(\alpha_s)\else{{O($\alpha_s$)}}\fi\chkspace}
\def\oalpsq{\relax\ifmmode O(\alpha_s^2)
           \else{{O($\alpha_s^2$)}}\fi\chkspace}
\def\oalpc{\relax\ifmmode O(\alpha_s^3)
           \else{{O($\alpha_s^3$)}}\fi\chkspace}
\def\oalpf{\relax\ifmmode O(\alpha_s^4)
           \else{{O($\alpha_s^4$)}}\fi\chkspace}

\def\rb{\relax\ifmmode R_3^b/R_3^{all}
           \else{{$R_3^b/R_3^{all}$}}\fi\chkspace}
\def\rc{\relax\ifmmode R_3^c/R_3^{all}
           \else{{$R_3^c/R_3^{all}$}}\fi\chkspace}
\def\ruds{\relax\ifmmode R_3^{uds}/R_3^{all}
           \else{{$R_3^{uds}/R_3^{all}$}}\fi\chkspace}
\def\ri{\relax\ifmmode R_3^i/R_3^{all}
           \else{{$R_3^i/R_3^{all}$}}\fi\chkspace}
\def\rj{\relax\ifmmode R_3^j/R_3^{all}
           \else{{$R_3^j/R_3^{all}$}}\fi\chkspace}
\def\alpi{\relax\ifmmode \alpha^i_s/\alpha^{all}_s
           \else{{$\alpha^i_s/\alpha^{all}_s$}}\fi\chkspace}

\def\mbz{\relax\ifmmode m_b(M_Z)
           \else{{$m_b(M_Z)$}}\fi\chkspace}
\def\mbb{\relax\ifmmode m_b(M_b)
           \else{{$m_b(M_b)$}}\fi\chkspace}

\def\plb{Phys. Lett.\chkspace}

\def\prl{Phys. Rev. Lett.\chkspace}
\def\prd{Phys. Rev.\chkspace}

\def\z0{{$Z^0$}\chkspace}
\def\z0{\relax\ifmmode Z^0 \else {$Z^0$} \fi\chkspace}
\def\Dst{\relax\ifmmode {\rm D}^* \else {D$^*$}\fi\chkspace}
\def\Dpl{\relax\ifmmode {\rm D}^+ \else {D$^+$}\fi\chkspace}
\def\D0{\relax\ifmmode {\rm D}^0 \else {D$^0$}\fi\chkspace}
\def\Kst{\relax\ifmmode {\rm K}^* \else {K$^*$}\fi\chkspace}
\def\K0{\relax\ifmmode {\rm K}^0_s \else {K$^0_s$}\fi\chkspace}
\def\Kpl{\relax\ifmmode {\rm K}^+ \else {K$^+$}\fi\chkspace}
\def\Kstz{\relax\ifmmode {\rm K}^{*0} \else {K$^{*0}$}\fi\chkspace}

\def\beq{\begin{equation}}
\def\eeq{\end{equation}}
\def\bea{\begin{eqnarray}}
\def\eea{\end{eqnarray}}

\begin{document}

\hfill{OUNP-99-11}

\vskip .5truecm

\hfill{August 5 1999}

\vskip 1.5truecm

\centerline{\large \bf
QCD AND $\gamma\gamma$ INTERACTIONS}
\centerline{\large\bf AT A HIGH-ENERGY LINEAR e$^+$e$^-$ COLLIDER}

\vskip 2truecm

\centerline{\large P.N. BURROWS$^*$}

\vskip 1truecm

\centerline{Particle \& Nuclear Physics,}
\centerline{Keble Rd., Oxford, OX1 3RH,}
\centerline{United Kingdom}

\vskip 2truecm

\centerline{\bf Abstract}

\vskip .5truecm

\noindent
A summary is presented of the key strong-interaction measurements that
could be made at a high-energy, high-luminosity \ep collider.

\vskip 2truecm

\centerline{\it Summary Talk Presented at}
\centerline{\it The Worldwide Study on Physics and Experiments}
\centerline{\it with Future Linear \ep Colliders,}
\centerline{\it Sitges, Spain, April 28 - May 5 1999}

\vfill
\noindent
{$*$ Supported by the UK Particle Physics \& Astronomy Research Council}\\
email: {\it p.burrows1@physics.ox.ac.uk}

\eject

\section{Introduction}

Strong-interaction measurements at a future high-energy linear e$^+$e$^-$
collider (LC) will form an important component of the physics 
programme. A 1 TeV collider has an energy reach comparable with
the LHC, and offers the possibility of testing QCD 
in the experimentally clean, more theoretically accessible
\ep environment. In addition, $\gamma\gamma$ interactions
will be delivered free by Nature, and a dedicated $\gamma\gamma$
collider is an additional option, allowing detailed measurements of
the relatively poorly understood photon structure.
Here I review the main topics; more details can be found in~\cite{previous}:

\noindent
$\bullet$
Precise determination of the strong coupling \alp.

\noindent
$\bullet$
Measurement of the $Q^2$ evolution of \alp, searches for
new coloured particles and constraints on the GUT scale.

\noindent
$\bullet$
Measurements of the \tt(g) system

\noindent
$\bullet$
Measurement of the total $\gamma\gamma$ cross section and 
the photon structure function.

\noindent
Related top-quark, $\gamma\gamma$ and theoretical topics are summarised 
elsewhere~\cite{topexp,toptheor,albert}. 

\section{Precise Determination of \alp}

The current precision of individual \alpmzsq
measurements is limited at best to several per cent~\cite{alphasrev}.
Since the uncertainty on \alp translates directly into an uncertainty 
on perturbative QCD (pQCD) predictions, especially for high-order multijet 
processes, it would be desirable to achieve much better precision. 
In addition, since the weak and electromagnetic couplings are known
with much greater precision, the error on \alp represents the
dominant uncertainty on our `prediction' of the scale for
grand unification of the strong, weak and electromagnetic forces~\cite{gut}. 

Several techniques for \alp determination are available
at the LC:

\subsection{Event Shape Observables}

The determination of \alp from event `shape' observables that are
sensitive to the 3-jet nature of the particle flow has been
pursued for 2 decades and is generally well understood~\cite{philalp}. 
In this method one usually forms a differential distribution, 
makes corrections for detector and hadronisation effects,
and fits a pQCD prediction to the data, allowing \alp to vary.
Examples of such observables are the event thrust and jet masses.

The latest generation of such \alp measurements, from SLC and LEP, has
shown that statistical errors below the 1\% level can be obtained
with samples of a few tens of thousands of hadronic events.
With the current LC design luminosities of 
$5\times10^{33}$/cm$^2$/s (NLC/JLC) and 
$3\times10^{34}$/cm$^2$/s (TESLA), at $Q$ = 500 GeV, 
tens/hundreds of thousands of \ep \ra \qq events would be produced each year,
and a statistical error on \alp below the 1\% level could
be achieved easily. 

Detector systematic errors, which relate mainly to
uncertainties on the corrections made for acceptance and resolution
effects, are under control at the 1-3\% level (depending 
on the observable). If the LC detectors are designed to
be very hermetic, with good tracking resolution and efficiency,
as well as good calorimetric
jet energy resolution, all of which are required for the search
for new physics processes, it seems reasonable to 
expect that the detector-related uncertainties can be beaten down to
the 1\% level or better.

\ep \ra \z0\z0, $W^+W^-$, or \tt events 
will present significant backgrounds to \qq events for QCD studies, and the
selection of a highly pure \qq event sample will not be 
as straightforward as at the \z0 resonance. The application of kinematic cuts 
would cause a significant bias to the event-shape
distributions, necessitating compensating corrections at the level of
25\%~\cite{bias}. 
More recent studies have shown~\cite{schumm} that the majority of
$W^+W^-$ events can
be excluded without bias by using only right-handed electron-beam
produced events in the \alp analysis. Furthermore, the application of
highly-efficient $b$-jet tagging can be used to reduce the \tt
contamination to the 1\% level. After statistical subtraction of the
remaining backgrounds (the \z0\z0 and $W^+W^-$ event properties (will)
have been measured accurately at SLC and LEP),
the residual bias on the event-shape distributions is expected to
be under control at the 1\% level on \alp.

Additional corrections must be made for the effects of the smearing of the
particle momentum flow caused by hadronisation.
These are traditionally evaluated using Monte Carlo models.
The models have been well tuned at SLC and LEP and are widely 
used for evaluating systematic
effects. The size of the correction factor, and hence the uncertainty, 
is observable dependent, but the `best' observables have 
uncertainties as low as 1\% on \alp. Furthermore, one expects
the size of these hadronisation effects
to diminish with c.m. energy at least as fast as 1/$Q$.
Hence 10\%-level corrections at the \z0 should dwindle to 
less than 2\% corrections at $Q$ $\geq$ 500 GeV, and the
associated uncertainties should be well below the 1\% level
on \alp.

Currently pQCD calculations of event shapes 
are available complete only
up to \oalpsq. Since the data contain knowledge of all
orders one must estimate the possible bias inherent in measuring
\alpmzsq using the truncated QCD series.
Though not universally accepted, it is customary to estimate this from
the dependence of the fitted \alpmzsq on the QCD renormalisation scale,
yielding a large and dominant uncertainty of about $\pm 0.007$~\cite{sldalp}.
Since the missing terms are \oalpc, and since 
\alp(500 GeV) is expected to be about 25\% smaller than \alpmzsq, 
one expects the uncalculated contributions to be
almost a factor of two smaller at the higher energy, leading to an
estimated uncertainty of $\pm0.004$ on \alp(500 GeV). However,
translating to the conventional yardstick \alpmzsq yields an uncertainty of
$\pm0.006$, only slightly smaller than currently.
Therefore, a 1\%-level \alpmzsq measurement is possible experimentally,
but will not be realised unless \oalpc contributions are
calculated.
 
\subsection{The \tt System}

The value of \alp controls the strong potential that binds quarkonia resonances. 
In the case of
\tt production near threshold, the large top mass and 
decay width ensure that the top quarks decay in a time comparable with
the classical period of rotation of the bound system, washing out most
of the resonant structure in the cross-section, $\sigma_{\tt}$. The shape of
$\sigma_{\tt}$ near threshold
hence depends strongly on both $m_t$ and \alp.
Fits of next-to-leading-order (NLO) pQCD calculations to 
simulated measurements of $\sigma_{\tt}$
showed~\cite{topalp} that $m_t$ is strongly correlated with \alp. 
Fixing \alp allowed the error on $m_t$ to be reduced 
by a factor of 2. Since the main aim of such an exercise is to
determine $m_t$ as precisely as possible, the optimal strategy
would be to {\it input} \alp from elsewhere. 

Moreover, recent NNLO calculations of $\sigma_{\tt}$ near threshold
have caused consternation, in that the size of the NNLO contributions
appears to be comparable with that of the NLO contributions, and the
change in the shape causes a shift of roughly 1 GeV in the value of the 
fitted $m_t$. This mass shift can be avoided by a judicious
top-mass definition~\cite{toptheor}, which also reduces the $m_t$-\alp correlation.
However, the resulting cross-section normalisation uncertainty translates into
an uncertainty of $\pm0.012$ on \alpmzsq, \ie about 5 times
larger than the estimated statistical error~\cite{topexp}. Although this may provide
a useful `sanity check' of \alp in the \tt system,
it does not appear to offer the prospect of a 1\%-level measurement.
 
A preliminary study has also been made~\cite{werner} of the determination
of \alp from $R$ = $\sigma_{\tt}/\sigma_{\mu^+\mu^-}$ {\it above} threshold.
For $Q$ $\sim$ 400 GeV the theoretical uncertainty on $R$ is roughly 3\%;
for $Q$ $\geq$ 500 GeV the exact value of $m_t$ is much less important and
the uncertainty is smaller, around 0.5\%. However, on the experimental
side the limiting precision on $R$ will be given by the uncertainty on
the luminosity measurement. If this is only as good as at LEPII, \ie around
2\%, then \alpmzsq could be determined with an experimental precision of
at best 0.007, which is not especially useful other than as a consistency
check.

Finally, there remains the possibility of determining \alp using
\ttg events, which have recently been
calculated~\cite{arnd} at NLO. For reasonable values of the
jet-resolution scale $y_c$ the NLO contributions are substantial, of
order 30\%, which is comparable with the situation for massless quarks.
The discussion of unknown higher-order contributions above is hence
also valid here, and \ttg events will only be 
useful for determination of \alp once the NNLO contributions have been calculated.
If the \ttg event rate can be measured precisely, the ansatz of
flavour-independence of strong interactions can be tested for the top quark,
and the running of $m_t$ could be determined in a similar manner to
the running $b$-quark mass~\cite{bmass}. 
A precision of 1\% implies a measurement of $m_t(Q)$ with an error of 5 GeV.

\subsection{A High-luminosity Run at the \z0 Resonance}

A LC run at the \z0 resonance is attractive for 
a number of reasons. At nominal design luminosity tens of millions of
\z0/day would be delivered, offering the possibility of
a year-long run to collect a Giga \z0 sample for ultra-precise
electroweak measurements and tests of radiative corrections.
Even substantially lower luminosity, or a shorter run, at the \z0
could be useful for detector calibration.

A Giga \z0 sample offers two additional options for 
\alp determination via measurements of the inclusive ratios 
$\Gamma^{had}_Z/\Gamma_Z^{lept}$ and
$\Gamma^{had}_{\tau}/\Gamma_{\tau}^{lept}$. Both are indirectly 
proportional to \alp, and hence require a very large event sample for 
a precise measurement. For example, the current LEP data sample of 16M
\z0 yields an error of 0.003 on \alpmzsq from 
$\Gamma^{had}_Z/\Gamma_Z^{lept}$. The statistical error could,
naively, be pushed to below the 0.0005 level, but systematic errors
arising from the lepton selection will probably limit the precision
to 0.0016~\cite{klaus}. Nevertheless this would be a very precise, reliable 
measurement. In the case of 
$\Gamma^{had}_{\tau}/\Gamma_{\tau}^{lept}$ the experimental precision
from LEP and CLEO is already at the 0.001 level on \alpmzsq. However,
there has been considerable debate about the size of the theoretical
uncertainties, with estimates ranging from 0.002 to 0.006. If this
situation is clarified, and the theoretical uncertainty is small,  
$\Gamma^{had}_{\tau}/\Gamma_{\tau}^{lept}$ may offer a further
1\%-level \alpmzsq measurement.

\section{$Q^2$ Evolution of \alp}

The running coupling is sensitive to the presence of any new
coloured particles, such as gluinos, beneath the c.m. energy threshold 
via their vacuum polarisation contributions. 
Measurements of event shape observables at high energies, combined
with existing lower energy data, would allow one to 
search for anomalous running.
In addition, extrapolation of the running \alp can be combined with extrapolations of the
dimensionless weak and electromagnetic couplings in order to
try to constrain the GUT scale~\cite{gut}. 
The highest-energy measurements, up to $Q$ = 200 GeV, are currently provided 
by LEPII. Older data from \epa span the range
$14\leq Q\leq 91$ GeV. A 0.5 - 1.0 TeV linear collider would increase 
significantly the lever-arm for measuring the running~\cite{previous,bias}.

However, over a decade from now the combination of LC data
with the older data may not be straightforward,
and will certainly not be optimal since some of the systematic
errors are correlated among data at different energies. 
It would be desirable to measure {\it in the same apparatus,
with the same technique, and by applying the same treatment to the
data} at least one low-energy point - at the \z0 or even lower - 
in addition to points at the $W^+W^-$ and \tt thresholds, as well
as at the highest c.m. energies.
 
\section{Other \ep QCD Topics}
 
Limited space allows only a brief mention
of several important topics~\cite{previous}:

\noindent
$\bullet$
Searches for anomalous chromo-electric and
chromo-magnetic dipole moments of quarks,
which effectively modify the rate and pattern of gluon
radiation.
Limits on the anomalous $b$-quark chromomagnetic moment have been
obtained at the \z0 resonance~\cite{sldchromo}.
The \ttg system would be important to study at the LC.

\noindent
$\bullet$
Gluon radiation in \tt events is expected to be strongly regulated by
the large mass and width of the top quark. Measurements of
gluon radiation patterns in \ttg events may provide 
additional constraints on the top decay width~\cite{lynne}.

\noindent
$\bullet$
Polarised electron (and positron) beams can be exploited 
to test symmetries using multi-jet final states.
For polarized \ep annihilation to
three hadronic jets one can define $\vec{S_e}\cdot(\vec{k_1}\times \vec{k_2})$,
which correlates the electron-beam polarization vector $\vec{S_e}$
with the normal to the three-jet plane defined by
$\vec{k_1}$ and $\vec{k_2}$, the momenta of the two quark jets.
If the jets are ordered by momentum (flavour)
the triple-product is CP even (odd) and T odd.
Standard Model T-odd contributions of this form are
expected~\cite{lance} to be immeasurably small, and limits 
have been set for the \bbg system~\cite{sldtodd}.
At the LC these observables will provide
a search-ground for anomalous effects in the \ttg system.

\noindent
$\bullet$
The difference between the particle multiplicity in heavy- ($b,c$)
and light-quark events is predicted~\cite{doksh} to be independent of
c.m. energy. Precise measurements have been made at the \z0, but 
measurements at other energies are limited in precision, rendering a
limited test of this important prediction. High-precision
measurements at the LC would add the lever-arm for a powerful test.

\noindent
$\bullet$
Colour reconnection and Bose-Einstein correlations are
fascinating effects. They are important to study precisely since
they may affect the precision with which the masses of heavy particles,
such as the $W^{\pm}$ and top-quark, can be reconstructed kinematically
via their multijet decays~\cite{torbj}.

\section{Photon Structure}

Though much progress has been made in recent years at LEP and HERA,
a thorough understanding of the `structure'  of the venerable photon
is still lacking. Away from the \z0 resonance the relative cross-section 
for $\gamma\gamma$ scattering is large, but good detector acceptance in
the low-polar-angle regions is required.
The LC provides an opportunity to make definitive
measurements, either from the `free' $\gamma\gamma$ events provided
in the \ep collision mode, or via a dedicated high-luminosity 
`Compton collider' facility. 
From the range of interesting $\gamma\gamma$
topics~\cite{albert} I mention only a few important `QCD' measurements:

\noindent
$\bullet$
The total cross-section, $\sigma_{\gamma\gamma}$, and the form of its rise
with $Q$, will place constraints on models which cannot be 
differentiated with today's data; `proton-like' models
predict a soft rise, whereas `minijet' models predict a steep rise.

\noindent
$\bullet$
The photon structure function, $F_2^{\gamma\gamma}(x,Q^2)$, and the
nature of its rise at low $x$ in relation to `BFKL' or `DGLAP'
evolution.

\noindent
$\bullet$
Polarised structure functions, the charm
content of the photon, and diffractive phenomena.

\section{Summary and Conclusions}

Tests of QCD will enrich the physics programme at a high-energy
\ep collider. Measurement of \alpmzsq at the 1\% level of
precision appears feasible experimentally, but will require
considerable theoretical effort.
A search for anomalous running of \alp($Q^2$) is an attractive
prospect, but presents serious requirements on the design of both
the collider and detectors. Electron-beam polarisation can
be exploited to perform symmetry tests using multi-jet final states.
Interesting gluon
radiation patterns in \tt events could be used to constrain the
top quark decay width. Measurement of the gluon radiation spectrum
would also constrain anomalous strong top-quark couplings.
Realistic hadron-level Monte Carlo simulations, including detector
effects, need to be performed to evaluate these possibilities
quantitatively. 

\section*{Acknowledgements}
I thank A. Brandenburg and A. de Roeck for their help in
preparing this summary



\begin{thebibliography}{99}

\bibitem{previous} 
P.N. Burrows, 
Proc. Workshop on
Physics and Experiments with Linear Colliders, Sept. 8-12 1995,
Morioka-Appi, Japan, World Scientific 1996,
Ed. A. Miyamoto \etal, p. 179.

\bibitem{topexp}
M. Martinez, top-quark (experiment) summary, these proceedings.

\bibitem{toptheor}
A. Hoang, top-quark (theory) summary, these proceedings.

\bibitem{albert} A. de Roeck, $\gamma\gamma$ and e$^-$e$^-$ summary, these proceedings.
 
\bibitem{alphasrev} 
P.N. Burrows, 
Proc. 3rd International Symposium on Radiative Corrections,
Cracow, Poland, August 1-5 1996; Acta Phys. Pol. {\bf B28} (1997) 467.\\
S. Bethke, 
PITHA-98-43;
to appear in Proc. 4th International Symposium on Radiative Corrections, 
Barcelona, Spain, 8-12 September 1998. 

\bibitem{gut} U. Amaldi \etal, Phys. Lett. {\bf B281} (1992) 374.
 
\bibitem{philalp} 
See \eg P.N. Burrows, 
Proc. XXVIII International Conference on High
Energy Physics, Warsaw, Poland, July 25-31 1996, Eds. Z.~Adjuk,
A.K.~Wroblewski, World Scientific 1997, p. 797.
 
\bibitem{bias}
S. Bethke, Proc. Workshop on Physics and Experiments with Linear
e$^+$e$^-$
Colliders, 26-30 April 1993, Waikoloa, Hawaii; World Scientific,
Eds. F.A. Harris \etal, Vol. II p. 687.
 
\bibitem{schumm}
B. Schumm, these proceedings.

\bibitem{sldalp}
See \eg SLD Collab., K. Abe \etal, \prd {\bf D51} (1995) 962.
 
\bibitem{topalp}
JLC Group, S. Matsumoto \etal, KEK Report 92-16 (1992) p. 53.\\
P. Comas \etal, Proc. Workshop on
Physics and Experiments with Linear Colliders, Sept. 8-12 1995,
Morioka-Appi, Japan, 
p. 455.
 
\bibitem{werner}
W. Bernreuther, talk at ECFA/DESY Workshop, Oxford, March 20-23 1999.

\bibitem{arnd}
A. Brandenburg, these proceedings.

\bibitem{bmass}
DELPHI Collab., P.~Abreu \etal, Phys.  Lett. {\bf B418} (1998) 430.\\
A. Brandenburg \etal, SLAC-PUB-7915 (1999), subm. to \plb B.

\bibitem{klaus}
K. M\"oenig, these proceedings.

\bibitem{sldchromo}
SLD Collab., K. Abe \etal, SLAC-PUB-8155 (1999).
 
\bibitem{lynne}
L. Orr, these proceedings.

\bibitem{lance}
A. Brandenburg, L. Dixon, Y. Shadmi, Phys. Rev. {\bf D53} (1996) 1264.
 
\bibitem{sldtodd}
SLD Collab., K. Abe \etal, SLAC-PUB-8156 (1999).
 
\bibitem{doksh}
Y. Dokshitzer \etal, \prl {\bf 69} (1992) 3025.
 
\bibitem{torbj}
T. Sj\"ostrand, these proceedings.
 
\end{thebibliography}
\end{document}